\documentclass{article}
\usepackage[margin=1in]{geometry}
\usepackage{amsfonts,amsmath,amssymb}
\usepackage[none]{hyphenat}
\usepackage{graphicx}
\usepackage{hyperref}
\usepackage{caption}
\usepackage{fullpage}
\usepackage{amsthm}
\usepackage{amscd}
\usepackage{indentfirst}
\usepackage{pict2e}  
\usepackage{epic}
\numberwithin{equation}{section}
\usepackage{epstopdf}
\usepackage{subcaption}
\usepackage{authblk}
\usepackage{float}
\usepackage{natbib}
\usepackage{xcolor}
\usepackage [autostyle]{csquotes}
\usepackage{chngcntr}
\counterwithout{equation}{section}

\begin{document}
\title{Rational Finance Approach to Behavioral Option Pricing}
\author[a]{Jiexin Dai}
\author[a]{Abootaleb Shirvani}
\author[b]{Frank J. Fabozzi}
\affil[a]{The Department of Mathematics and Statistics, Texas Tech University}
\affil[b]{EDHEC Business School}
\date{}

\maketitle

\noindent \textbf{Abstract:}   
When pricing options, there may be different views on the instantaneous mean return of the underlying price process. According to \citet{black1972capital}, where there exist heterogeneous views on the instantaneous mean return, this will result in arbitrage opportunities. Behavioral finance proponents argue that such heterogenous views are likely to occur and this will not impact option pricing models proposed by rational dynamic asset pricing theory and will not give rise to volatility smiles. To rectify this, a leading advocate of behavioral finance has proposed a behavioral option pricing model.  As there may be unexplored links between the behavioral and rational approaches to option pricing, in this paper we revisit \citet{shefrin2008behavioral} option pricing model as an example and suggest one approach to modify this behavioral finance option pricing formula to be consistent with rational dynamic asset pricing theory by introducing arbitrage transaction costs which offset the gains from arbitrage trades.\\


\section{Introduction}
\indent Proponents of behavioral finance have identified several market anomalies (i.e., empirical findings that are inconsistent with theories formulated by traditional finance). Since the late 1970s (see \citet{heukelom2014behavioral}) there have been numerous papers that seek to explain these anomalies by building upon the foundational ideas of behavioral economics proposed by Daniel Kahneman and Amos Tversky. The debate regarding whether financial market agents are rational or irrational in making financial decisions is ongoing\footnote{See, for example, \citet{zeckhauser1986comments}, \citet{hirshleifer2001investor}, \citet{shiller2003efficient}, \citet{barberis2003survey}, \citet{brav2004rational}, \cite{curtis2004modern}, \cite{parisi2005law} in Chapter 21, and \cite{thaler2005advances}.}. Mark Rubinstein, an advocate of rational markets, criticized the behavioral finance approach to asset pricing. \citet{rubinstein2001rational} argued that as a trained financial economist, he was taught that the “Prime Directive” in pricing is to explain asset prices by using rational models. This is not to say that pricing cannot rely on irrational investor behavior if rational models fail to price assets properly (i.e., fail to correctly price market prices).  In his opinion, the behavioralist literature “has lost all the constraints of this directive”. Meir Statman, one of the leading academics who has contributed to the behavioral finance camp, has a different perspective. \citet{statman1995behavioral} argues that “standard finance is indeed so weighted down with anomalies that it makes much sense to continue the reconstruction of financial theory on behavioral lines.”\\
\indent Proponents of the behavioral finance did not merely critique rational finance models on asset pricing, option pricing, and portfolio selection. They have proposed behavioral finance based on irrational behavior.  Meir Statman and Hersh Shefrin have taken the lead on formulating such models. For example, \cite{shefrin1993behavioral} first proposed option pricing model based on prospect theory. \citet{shefrin1999irrational} argued that “we live in a world of heterogeneous beliefs” and “option markets are particular vulnerable in this respect”.  Later, \citet{shefrin2008behavioral} proposed an equilibrium approach to option pricing in which the representative agents view the return from the underlying asset as a mixture of two different normal distributed returns representing the heterogeneous views on the asset return of the buyer and the seller of the option. More specifically, Shefrin defines a market model with two investors sharing two price processes with common Brownian motion as market driver, the same volatility parameters and different instantaneous mean returns. Since the mixture of two different log-normal distributions is not infinitely divisible\footnote{See \citet{steutel2003infinite}}, the price process of the representative agent is not a semi-martingale, which according to Black (1972), Shefrin’s proposed model allows for arbitrage opportunities. \citet{rockenbach2004behavioral} reports the arbitrage-free option pricing is invalidated based on mental accounts, which result is consistent with behavioral portfolio theory of Shefrin and Statman (see \citet{shefrin2000behavioral}). \citet{pena2010role} derived a behavioral Black-Scholes option pricing model but the underlying price process is not a semi-martingale, thus, according to the fundamental theorem of asset pricing, this model leads to arbitrage (see \cite{delbaen1994general}). \citet{barberis2019prospect} incorporate prospect theory into asset pricing models trying to explain market anomalies, they modified expected utility from the cumulative prospective individual through value function and probability weighting function. However, after transformations under the value function and probability weighting function, the underlying price process is no longer infinitely divisible and for that reason the prospect theory based asset pricing models lead to arbitrages. \\
\indent Within Rational Dynamic Asset Pricing Theory (RDAPT) the most important problem is the characterization of economically rational consistent models for financial markets (see \citet{duffie2010dynamic}). In RDATP, the central assumption is that there is no-arbitrage: a market participant should not engage in a contract in which the market participant can lose an infinite amount of money in a frictionless market. Regardless of how irrational a representative agent might be, one should not be so irrational as to be subject to an infinite loss, and the assumption of infinitely divisibility is crucial to no-arbitrage option pricing (see \citet{bayraktar2016arbitrage}). {\citet{brav2004rational} mentioned “perhaps every equilibrium prediction that assumes the survival of the irrationality may require a (shadow) prediction from some rational model”. Indeed, one should incorporate human beings' behavior to the modeling the dynamic of asset prices. However, as \citet{miller1986behavioral} pointed out that for individual investors\footnote{See \citet{brav2004rational}: “individual investors who hold modest amounts of stock directly and who, unlike institutional and other large investors, do not rely heavily on professional portfolio advisers”.} unlike those institutional investors, there may be numerous nontrivial life-related concerns associated with each trading activity. But the purpose of proposing financial models is neither paying too much attention on human behaviors to get lost in millions of details nor completely ignore human behaviors. Rather, focusing on “pervasive market forces” is the direction of proposing abstract asset pricing model. We are proposing abstract asset pricing models to incorporating human behaviors but not forgetting our “principal concern”. No-arbitrage, as our “principal concern”, will drive any mispricing caused by irrational traders' bad investment strategy to zero. There is no reason for behavioralists to object to the no-arbitrage assumption which is as a fundamental notion for both the rational or behavioral finance camps.  If this assumption is not satisfied, agents using financial asset pricing formulas allowing for arbitrage could suffer tremendous losses. If a trader applies a behavioral option pricing model such as Shefrin (2008) by being long in the contract, there will be a “rational” trader who will take the short position and apply arbitrage strategy. Indeed, in Shefrin's option pricing approach there is no proposed hedging strategy the option seller (i.e., the short) could use and the reason is that there is no such strategy possible. \\
\indent In this paper, we use Shefrin's option pricing model as an example, and suggest one approach to adjust this option pricing formulas for option traders having heterogeneous views on the underlying pricing process so that those formulas are consistent with the RDAPT. Specifically, we impose trading costs (so-called arb-costs) which will offset the gains from possible arbitrage opportunities in the market. Equilibrium option pricing models when traders have heterogeneous beliefs have been studied in the RDART literature. Our approach to option pricing in the presence of heterogeneous beliefs can be roughly explained as follows: for the hedger to realize arbitrage profits, the hedger must be able to trade at a high speed, imposing arb-costs on the velocity of trades can offset the gains the hedger accumulates when applying the arbitrage trade. In standard hedging when no arbitrage occurs, the arb-costs are not significant. \\
\indent We have organized the paper as follows. In Section 2 we identify the flaw in Shefrin's behavioral approach to option pricing. In Section 3 we introduce an arb-cost on the binomial tree of Shefrin’s behavioral market model and derive the risk-neutral price dynamics as well as the resulting risk-free rate. Our numerical analysis of observed call option prices is based on SPDY S{\&}P 500 index (SPY) and iShare Core S{\&}P 500 (IVV). Our conclusion is summarized in Section 5.
\section{Shefrin's Behavioral Option Pricing Model}
\indent Shefrin constructed asset pricing model by defining a sole agent (the representative investor) as the representative of society as a whole, capable of aggregating the behavior of large numbers of economic agents. He then extended the asset pricing model that was constructed based on the representative investor who was not rational in order to include heterogeneous beliefs. Shefrin constructed an option pricing model by introducing two investors. Both investors agree on the risk-free asset process and volatility of the risky asset, but disagree on the drift term for the risky asset. Investor 1 believe that the stock price $S$ obeys the process 
\begin{equation}
\frac{d S}{S}=\mu_{1} d t+\sigma dZ
\end{equation}
where $Z$ is a Wiener process. Investor 2 believe that the stock price $S$ obeys the process
\begin{equation}
\frac{d S}{S}=\mu_{2} d t+\sigma dZ
\end{equation}
Shefrin claims the option will be priced according to Black–Scholes in this setting. Furthermore, he concludes that heterogeneity will neither impact option prices nor give rise to volatility smiles.\\
\indent In general, the above claim from \cite{shefrin2008behavioral} is not true. To see that, suppose investor 1 takes the long position in the European option contract $\mathbf{C}$ with: \\
\indent (1) price process $C(t)=f(S(t),t),t\geq 0$; \\
\indent (2) maturity time T;  and \\
\indent (3) payoff function at maturity $f(S(T),T)=g(T)$, where $f(x,t),x>0,t\geq 0$ is sufficiently smooth.\\
Then the dynamics of the long position in $\mathbf{C}$ is given by the Itô formula:
\begin{equation}
\begin{aligned} d C(t) &=d f(S(t), t) \\ &=\left(\frac{\partial f(S(t), t)}{\partial t}+\frac{\partial f(S(t), t)}{\partial x} \mu_{1} S(t)+\frac{1}{2} \frac{\partial^{2} f(S(t), t)}{\partial x^{2}} \sigma^{2} S(t)^{2}\right) d t+\frac{\partial f(S(t), t)}{\partial x} \sigma S(t) d Z(t) \end{aligned}
\end{equation}
Suppose investor 2 takes the short position in $\mathbf{C}$, and forms a self-financing strategy 
\begin{equation}
C(t)=f(S(t), t)=a(t) S(t)+b(t) \beta(t), t \geq 0
\end{equation} 
where $\beta(t)=\beta(0) e^{r t}, t \geq 0$ is the riskless asset with a riskless rate $r$. The the dynamic of the replicating portfolio is given by
\begin{equation}
\begin{aligned} d C(t) &=d f(S(t), t) \\ &=a(t) d S(t)+b(t) d \beta(t) \\ &=\left(a(t) \mu_{2} S(t)+b(t) r \beta(t)\right) d t+a(t) \sigma S(t) d Z(t) \end{aligned}
\end{equation}
Equating the expression for $d C(t)$ leads to $a(t)=\frac{\partial f(S(t), t)}{\partial x}$, and 
\begin{equation}
\frac{\partial f(S(t), t)}{\partial t}+\frac{\partial f(S(t), t)}{\partial x} \mu_{1} S(t)+\frac{1}{2} \frac{\partial^{2} f(S(t), t)}{\partial x^{2}} \sigma^{2} S(t)^{2}=a(t) \mu_{2} S(t)+r(f(S(t), t)-a(t) S(t))
\end{equation}
Setting $S(t)=x$, leads to the following partial differential equation (PDE): 
\begin{equation}
\frac{\partial f(x, t)}{\partial t}+\frac{\partial f(x, t)}{\partial x}\left(\mu_{1}-\mu_{2}+r\right) x-r f(x, t)+\frac{1}{2} \frac{\partial^{2} f(x, t)}{\partial x^{2}} \sigma^{2} x^{2}=0
\end{equation}
which is the Black-Scholes formula in the homogeneous case $\mu_{1}=\mu_{2}$. For $\mu_{1} \neq \mu_{2}$, the above PDE will lead to option pricing with arbitrage opportunities.

\section{Rational Option Pricing with Transaction Cost}
We now start with the description of Shefrin’s model, see \cite{shefrin2008behavioral}, Chapter 8, but from the viewpoint of RDAPT. Consider a financial market with  two investors sharing an aggregate consumption (AC) $\omega(0)>0$ amount at $t^{(0)}=0$. At any subsequent period $t^{(k+1)}=(k+1) \Delta t, k=0,1, \ldots, n-1, t^{(n)}=T<\infty, n \uparrow \infty$, the aggregate amount available will unfold through a binomial process, growing by $u^{(\Delta t)}>1$ or $d^{(\Delta t)}<1$. Under Shefrin's model assumption, the investor $\beth^{(j)}, j=1,2$ attaches probability $p^{(j)}(\Delta t)$ for an upward movement of the AC-process and $1-p^{(j)}(\Delta t)$ for a downturn movement; that is, the discrete AC-dynamics is given by 
\begin{equation}
\omega^{(j)}\left(t^{(k+1)}\right)=\left\{\begin{array}{c}{\omega^{(j)}\left(t^{(k)}\right) u^{(\Delta t)}, \; w \cdot p \cdot p^{(j)}(\Delta t)} \\ {\omega^{(j)}\left(t^{(k)}\right) d^{(\Delta t)},\; w \cdot p \cdot 1-p^{(j)}(\Delta t)}\end{array}\right.
\end{equation}
Following Shefrin (2005), Chapter 8, suppose trader $\beth^{(S)}$ (resp. $\beth^{(V)}$ trades a risky asset (stock) $\mathfrak{M}$ on a binomial lattice with price dynamic $S_{k \Delta t}, k \in \mathcal{N}^{(0)}, \; with \; S_{0}>0$ (resp. $V_{k \Delta t}, k \in \mathcal{N}^{(0)}, \; with \; V_{0}>0$). The joint price process dynamic is given by the following binomial tree \footnote{This binomial tree (11) was introduced in Kim at al (2016) (see also Jarrow and Rudd (2008)) as an extension of the classical CRR-model (\citet{cox1979option}). We use this more general binomial pricing tree, because we require the bivariate pricing tree to be driven by one risk factor, and with that requirement, CRR-model is not appropriate.}. 
\begin{equation}
\left[\begin{array}{c}{\left.S_{(k+1) \Delta t}\right.} \\ {V_{(k+1) \Delta t}}\end{array}\right]=\left\{\begin{array}{l}{\left[\begin{array}{l}{S_{(k+1) \Delta t: u p}=S_{k \Delta t}(1+\mu \Delta t+\sigma \sqrt{\Delta t})} \\ {V_{(k+1) \Delta t ; u p}=V_{k \Delta t}(1+m \Delta t+v \sqrt{\Delta t})}\end{array}\right] w \cdot p \cdot \frac{1}{2}} \\ {\left[\begin{array}{c}{S_{(k+1) \Delta t: d o w n}=S_{k \Delta t}(1+\mu \Delta t-\sigma \sqrt{\Delta t})} \\ {V_{(k+1) \Delta t: d o w n}=V_{k \Delta t}(1+m \Delta t-v \sqrt{\Delta t})}\end{array}\right] w \cdot p \cdot \frac{1}{2}}\end{array}\right.
\end{equation}
$k \in \mathcal{N}^{(0)}, \Delta t>0, \mu \in \mathcal{R}, m \in \mathcal{R}, \sigma>0, v>0$\footnote{Here, and in what follows, all terms of order $o(\Delta t)$ are assumed to be 0.}. For every fixed $T>0$, the the bivariate binomial tree $\left(S_{k \Delta t}, V_{k \Delta t}\right)_{k \in 0, \ldots, N \Delta t}$ generates a bivariate polygon process with trajectories in the Prokhorov space $C\left([0, T]^{2}\right)$ which converges weakly to the following bivariate geometric Brownian motion $\left(S_{t}, V_{t}\right)_{t \in[0, T]}$\footnote{The proof is similar to that in Davydov and Rotar (2008)\citet{kim2016multi}, Theorem 2, and thus is omitted.}:
\begin{equation}
S_{t}=S_{0} e^{\left(\mu-\frac{\sigma^{2}}{2}\right) t+\sigma B(t)}, V_{t}=V_{0} e^{\left(m-\frac{v^{2}}{2}\right) t+v B(t)}, t \in[0, T]
\end{equation}
where $B(t), t \geq 0$, is a Brownian motion generating a stochastic basis $\left(\Omega, \mathcal{F}, \mathbb{F}=\left(\mathcal{F}_{t}, t \geq 0\right), \mathbb{P}\right)$.\\
\indent As we pointed out in Section 2, the above model is not free of arbitrage. To rectify this model, we will introduce a new model with transaction costs\footnote{Transaction costs include commissions, execution costs and opportunity costs. The investment costs have a fixed component and a variable component\citet{collins1991methodology}.} to eliminate the possible gains from arbitrage trading in the above model.\\
\indent To this end, let us consider a perpetual European derivative contract $\mathcal{G}$. Following Shefrin's model, $\mathcal{G}$ has price process $G_{k \Delta t}=G\left(S_{k \Delta t}, V_{k \Delta t}\right), k \in \mathcal{N}^{(0)}$. Next, we shall derive the trading dynamics of a representative investor $\mathfrak{N}$ who is observing historical trading activities of $\beth^{(S)}$ and $\beth^{(V)}$. We assume $\mathfrak{N}$ (as a representative agent) has taken simultaneously both the long and the short position in $\mathcal{G}$. $\mathfrak{N}$ trades $S_{t}$ (resp. $V_(t)$) as $\beth^{(S)}$ (resp. $\beth^{(V)}$) would do. $\mathfrak{N}$ forms a self-financing strategy $\left(a_{k \Delta t}, b_{k \Delta t}\right), k \in \mathcal{N}$ that generating a self-financing portfolio $P(t)=a(t) S(t)+b(t) V(t)$. Thus, \begin{equation}
G\left(t^{(k)}\right)=g\left(s\left(t^{(k)}\right), \; V\left(t^{(k)}\right)\right)=P\left(t^{(k)}\right)=a\left(t^{(k)}\right) S\left(t^{(k)}\right)+b\left(t^{(k)}\right) V\left(t^{(k)}\right)
\end{equation}
when $\mathfrak{N}$ trade as $\beth^{(S)}$ (resp. $\beth^{(V)}$), at any time interval $\left[t^{(k)}, t^{(k+1)}\right)$ the trade is subject to transaction cost
\begin{equation}
\left(\frac{S\left(t^{(k+1)}\right)}{S\left(t^{(k)}\right)}\right)^{\rho(\mathcal{S})}\left(\operatorname{resp} \cdot\left(\frac{V\left(t^{(k+1)}\right)}{v\left(t^{(k)}\right)}\right)^{\rho(\mathcal{V})}\right)
\end{equation}
where $\rho(\mathcal{S})=\mathfrak{C} \frac{\mu}{\sigma}$ (resp. $\rho(\mathcal{V})=\mathfrak{C} \frac{\mu}{\sigma}$) and $\mathbb{C}$ is an absolute constant.\\
\indent Next, $\mathfrak{N}$ choose $\left(a\left(t^{(k)}\right), b\left(t^{(k)}\right)\right)$, so that
\begin{equation}
-g\left(S\left(t^{(k+1)}\right), V\left(t^{(k+1)}\right)\right)+a\left(t^{(k)}\right) S\left(t^{(k+1)}\right)\left(\frac{S\left(t^{(k+1)}\right)}{s\left(t^{(k)}\right)}\right)^{\rho(S)}+b\left(t^{(k)}\right) V\left(t^{(k+1)}\right)\left(\frac{V\left(t^{(k+1)}\right)}{v\left(t^{(k+k)}\right)}\right)^{\rho(v)}=0
\end{equation}
That is, at $t^{(k+1)}$ the hedged portfolio plus the short position in $\mathcal{G}$ has value zero for all states of the world, and thus its value at $t^{(k)}$ should also be zero. Thus,
\begin{equation}
g\left(S\left(t^{(k)}\right), V\left(t^{(k)}\right)\right)=P\left(t^{(k)}\right)=a\left(t^{(k)}\right) S\left(t^{(k)}\right)+b\left(t^{(k)}\right) V\left(t^{(k)}\right)
\end{equation}
With $\rho(\mathcal{S})=\mathfrak{C} \frac{\mu}{\sigma}$ (resp. $\rho(\mathcal{V})=\mathfrak{C} \frac{\mu}{\sigma}$), we obtain the binomial option price dynamics:
\begin{equation}
\begin{aligned} g\left(s\left(t^{(k)}\right), V\left(t^{(k)}\right)\right) &=\\ &=Q^{(\Delta t)} g\left(S\left(t^{(k+1, u p)}\right), V\left(t^{(k+1, u p)}\right)\right)+\left(1-Q^{(\Delta t)}\right) g\left(S\left(t^{(k+1, d o w n)}\right), V\left(t^{(k+1, d o w n)}\right)\right) \end{aligned}
\end{equation}
and the risk-neutral probabilities ($\mathfrak{N}$'s state-price probabilities) are $Q^{(\Delta t)}$ and $1-Q^{(\Delta t)}$, where
\begin{equation}
Q^{(\Delta t)}=\frac{1}{2}-\frac{\mu\left(1+\mathbb{C} \frac{\mu}{\sigma}\right)\left(1+\mathbb{C} \frac{\sigma}{2}\right)-m\left(1+\mathbb{C} \frac{m}{v}\right)\left(1+\mathbb{C} \frac{v}{2}\right)}{2(\sigma-v+\mathbb{C}(\mu-m))} \sqrt{\Delta t}
\end{equation}
Note that even if $\sigma=v$ (which is an arbitrage pricing model if $\beth^{(S)}$ and $\beth^{(V)}$ trades without arb-cost), as soon as $\mu \neq m$, risk-neutral probabilities exist, and thus the introduction of transaction costs $1+\rho(\mathcal{S}) \ln \frac{S\left(t^{(k+1)}\right)}{S\left(t^{(k)}\right)}$ (resp. $1+\rho(\mathcal{V}) \ln \frac{V\left(t^{(k+1)}\right)}{V\left(t^{(k)}\right)}$) has offset the arbitrage gains. From \citet{kim2016multi} Section 3.2, and \citet{black1972capital},  $Q^{(\Delta t)}$ should have the representation:
\begin{equation}
Q^{(\Delta t)}=\frac{1}{2}-\frac{1}{2} \frac{\mu^{(*)}-r^{(*)}}{\sigma^{(*)}} \sqrt{\Delta t}=\frac{1}{2}-\frac{1}{2} \frac{m^{(*)}-r^{(*)}}{v^{(*)}} \sqrt{\Delta t}
\end{equation}
where,
\begin{equation}
r^{(*)}:=\frac{\mu^{(*)} v^{(*)}-m^{(*)} \sigma^{(*)}}{v^{(*)}-\sigma^{(*)}}
\end{equation}
\begin{equation}
\mu^{(*)}:=\mu\left(1+\mathfrak{C} \frac{\mu}{\sigma}\right)\left(1+\mathfrak{C} \frac{\sigma}{2}\right)
\end{equation}
\begin{equation}
m^{(*)}:=m\left(1+\mathfrak{C} \frac{m}{v}\right)\left(1+\mathfrak{C}_{2}^{v}\right)
\end{equation}
\begin{equation}
\sigma^{(*)}:=\sigma+\mathfrak{C} \mu
\end{equation}
\begin{equation}
v^{(*)}:=v+\mathfrak{C} m
\end{equation}

$r^{(*)}$ is $\mathfrak{N}$'s risk-neutral rate, and $\mu^{(*)}$, $m^{(*)}$, $\sigma^{(*)}$, and $v^{(*)}$ are the adjusted (for arb-cost) drift and volatility parameters. Now the price process $\left(S_{t}, V_{t}\right)_{t \in[0, T]}$ as seen by $\mathfrak{N}$ in the risk-neutral world $\left(\Omega, \mathcal{F}, \mathbb{F}=\left(\mathcal{F}_{t}, t \geq 0\right), \mathbb{Q}\right)$, $\mathbb{Q}\:{\sim\mathbb{P}}$ has a dynamic given by
\begin{equation}
S_{t}=S_{0} e^{\left(r^{(*)}-\frac{\sigma^{(*)}}{2}\right) t+\sigma^{(*)} B^{(*)}(t)}, V_{t}=V_{0} e^{\left(r^{(*)}-\frac{v^{(*)}}{2}\right) t+v^{(*)} B^{(*)}(t)}, t \in[0, T]
\end{equation}
$B^{(*)}(t), t \geq 0$ is a Brownian motion on $\mathbb{Q}$, and an arithmetic Brownian motion on $\mathbb{P}$ with $B^{(*)}(t)=B(t)+\theta^{(*)} t$. The parameter $\theta^{(*)}=\frac{\mu^{(*)}-r^{(*)}}{\sigma^{(*)}}=\frac{m^{(*)}-r^{(*)}}{v^{(*)}}$ is the market price of risk in $\mathfrak{N}$'s market model with arb-costs. Note that the risk-neutral probability without arb-costs is 
\begin{equation}
Q^{(\Delta t ; n o ~ a r b-c o s t)}:=\frac{1}{2}-\frac{\mu-m}{2(\sigma-v)} \sqrt{\Delta t}
\end{equation}
Thus, $Q^{(\Delta t ; no ~ arb-cost)}$ is the risk-neutral probability in Black's (1972) model, and 
\begin{equation}
Q^{(\Delta t ; no ~ trans ~ cost)}:=\frac{1}{2}-\frac{1}{2} \frac{\mu-r}{\sigma} \sqrt{\Delta t}
\end{equation}
where $\frac{\mu-r}{\sigma}=\frac{m-r}{v}=\frac{\mu-m}{\sigma-v}$ and $r=\frac{\mu v-m \sigma}{v-\sigma}$. $\mathfrak{N}$'s model can be viewed as an extension of the Black (1972) model when our special type of transaction costs is introduced.
\section{Numerical example}
In this section, we apply the method introduced in Section 3 to a cross-sectional data analysis. Following the set-up in Shefrin's option pricing model: two representative spot traders have different views on one stock price process. We use the SPDR S{\&}P 500 ETF (SPY) and Vanguard S{\&}P 500 ETF (IVV) option prices as a data analysis example since both SPY and IVV track the Standard {\&} Poor 500 index which has been considered as a benchmark for the U.S. equity. We use the historical call option price for the SPY and IVV to calibrate the implied risk-free interest rate ($r^*$) and the implied volatility ($\sigma^*$) to calculate the optimized absolute constant $\mathfrak{C}$ as in equation 14. Then, the value for $r^*$ across time was calculated by using historical return data for SPY and IVV. Next, having the value of $r^*$ over time, we plot the option pricing implied volatility surfaces for SPY and IVV. Lastly, arbitrage cost ($\mathfrak{C}$) surfaces and the spread arbitrage cost coefficient surfaces for SPY and IVV were plotted in order to compare the liquidity spread surfaces. 
\subsection{Data}
The data used in this study are the daily stock price of SPDY S{\&}P 500 (SPY) and iShare Core S{\&}P 500 (IVV) obtained from Yahoo Finance from 09/09/2010 to 10/22/2019. The Treasury yield 10-years (TNX) is used as the riskless asset and the price data is obtained from Yahoo Finance. The European call option prices of SPY on 10/22/2019 with different time-to-maturity and strike prices were obtained from Chicago Board Options Exchange (CBOE)\footnote{see CBOE website http://www.cboe.com/delayedquote/quote-table-download}.
\subsection{The Calculation of $\mathfrak{C}$}
As mentioned in Section 3, we use equation 14 to calibrate $\mathfrak{C}$. The price process of SPY in risk-neutral world is 
\begin{equation}
S_{t}=S_{0} e^{\left(r^{(\cdot)}-\frac{\sigma^{(\cdot)^{2}}}{2}\right) t+\sigma^{(\cdot)} B^{(\cdot)}(t)}
\end{equation}
where $B^{(*)}(t), t \geq 0$ is a Brownian motion on $\mathbf{Q}$, and an arithmetic Brownian motion on $\mathbf{P}$ with $B^{(*)}(t)=B(t)+\theta^{(*)} t$\footnote{As mentioned in Section 3, the parameter $\theta^{(*)}=\frac{\mu^{(*)}-r^{(*)}}{\sigma^{(*)}}=\frac{m^{(*)}-r^{(*)}}{v^{(*)}}$ is the market price of risk in $\mathfrak{N}$'s market model with arb-costs.}. 

We use call option prices for the SPY and the price data is obtained from the CBOE on 10/22/2019 with different expiration dates and strike prices. The SPY stock price is used to calibrate the implied risk-free interest rate ($r^*$) and implied volatility ($\sigma^*$). The expiration date for the call option on SPY varies from 10/25/2019 to 01/21/2022, the prices per contract varies from ${\$}0.005$ to ${\$}274.105$ and the strike price varies from ${\$}25$ to ${\$}430$ among 3,330 different call option contracts. \\
\indent The SPY stock price as the underlying for the call option was ${\$}299.03$ quoted on CBOE on 10/22/2019. The sample mean and sample standard deviation for the SPY prices from 09/09/2010 to 10/22/2019 as the underlying on the call option were used as the estimated values for $\mu$ and $\sigma$ in equation 14. By using Black-Scholes-Merton model, the optimized implied risk-free interest rate ($r^*$) and implied volatility ($\sigma^*$) for the SPY call option price were calibrated and we obtained $r^{(*)}=0.0134$ and $\sigma^{(*)}=0.1385$, having the estimated $\sigma=0.0091$ and $\mu=0.0004321$\footnote{sample standard deviation and sample mean of daily returns of SPY from 09/09/2010 to 10/22/2019.} and the following equation (3.14):
\begin{equation}
\sigma^{(*)}:=\sigma+\mathfrak{C} \mu
\end{equation}
We obtained the optimized value of $\mathfrak{C}$ as a constant is 299.4773 and this value was used to calibrate the risk-free rate over time in the next subsection.
\subsection{Plot of the risk-free rate $r^{(*)}$}
\indent The risk-free rate $r^{(*)}$ is calibrated by using the historical stock return data of SPY and IVV as the underlying from 01/03/2011 to 10/22/2019. We calibrated the values of $\mu^{(*)}$, $m^{(*)}$, $\sigma^{(*)}$, and $v^{(*)}$ by applying equations (12), (13), and (14), respectively\footnote{$\mathfrak{C}$ as a constant is 299.4773 was applied in each equation.}, We use the “rolling method” to find daily $r^{(*)}$ with a fixed window size of 252 days. The smoothing plot of the risk-free rate is shown in Figure \ref{risk-free rate}. As mentioned in Section 3, this is representative investor's ($\mathfrak{N}$) risk-neutral rate.
\begin{figure}
	\includegraphics[width=\textwidth]{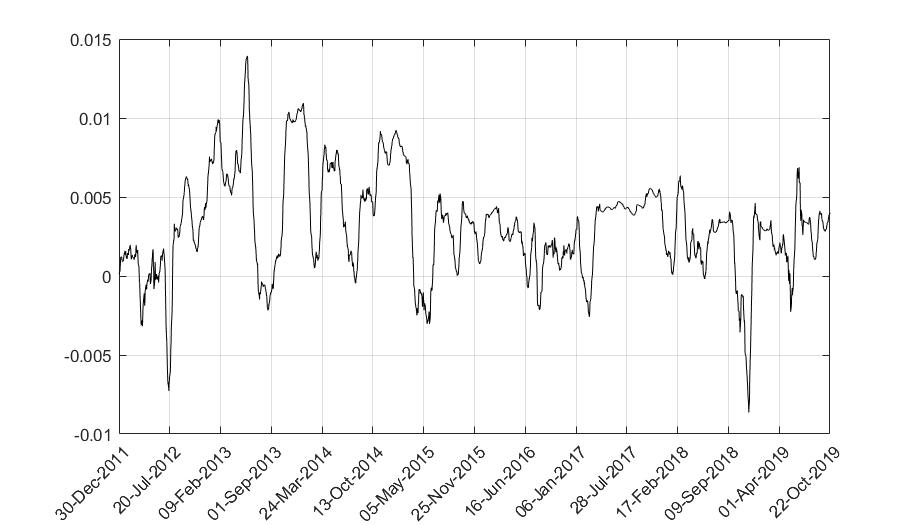} 
	\centering
	\caption{The plot of daily $r^{(*)}$ with window size per year}
	\label{risk-free rate}
\end{figure}
\subsection{Implied volatility surfaces and arbitrage cost surfaces}
By having the value of the implied risk-free rate $r(*)$ across time, we use option price data to plot the Black-Scholes implied volatility surface for various strike prices and maturities at a point in time for SPY and IVV, respectively, as shown in Figure \ref{IV for SPY} and Figure \ref{IV for IVV}. The implied volatility surfaces flatten out as the time to maturity increases for both SPY and IVV.\\
\indent In the plots of the arbitrage coefficient $\mathfrak{C}$ (ACS) surfaces as shown in Figure \ref{ACS for SPY} and Figure \ref{ACS for IVV}, we used the mid-point of the bid and ask prices for the SPY and IVV. \\
\begin{figure}
	\centering
		\begin{subfigure}{0.5\textwidth}
			\includegraphics[width=\textwidth]{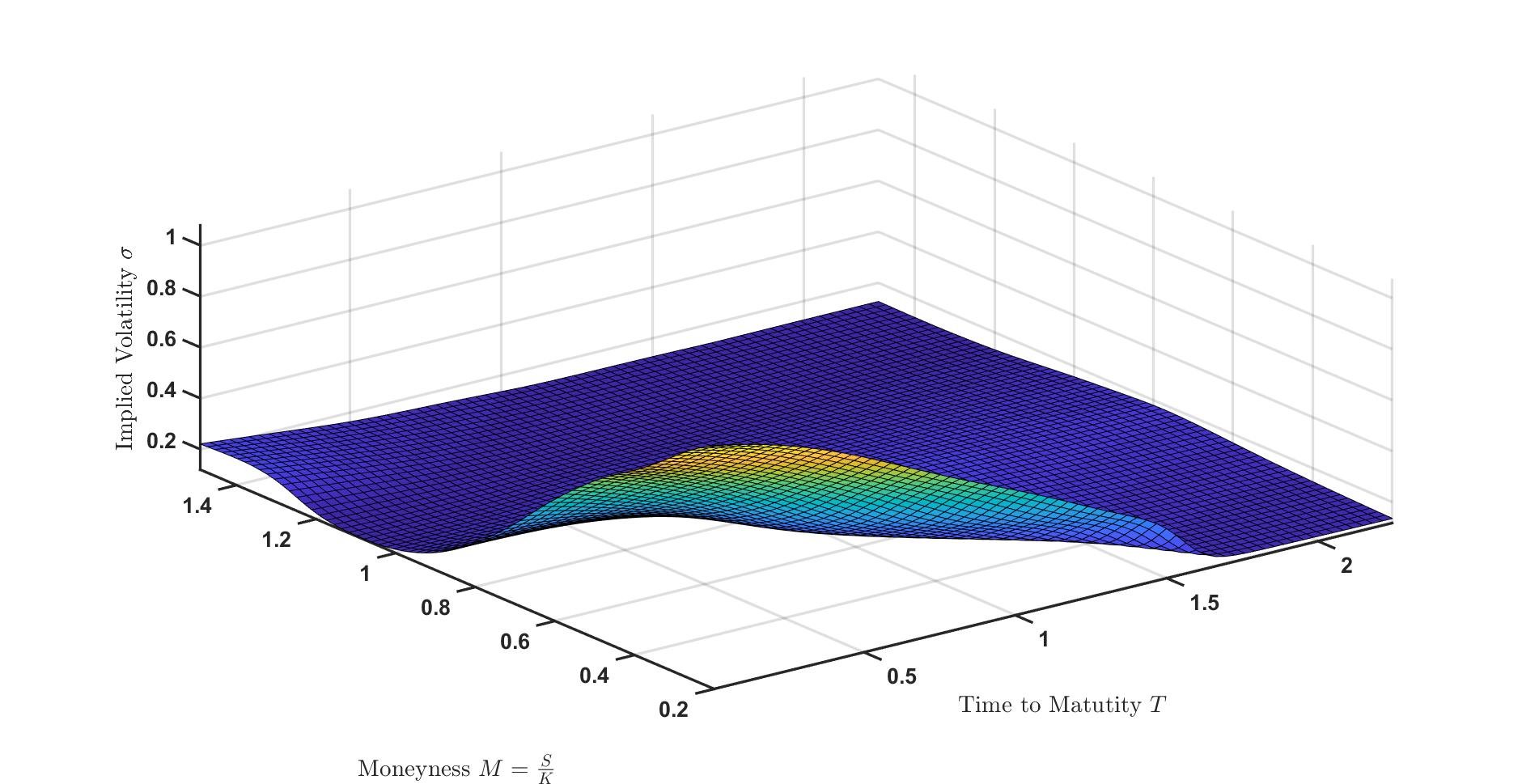}
			\caption{The plot of implied volatility for SPY}
			\label{IV for SPY}
		\end{subfigure}
		~
		\begin{subfigure}{0.5\textwidth}
			\includegraphics[width=\textwidth]{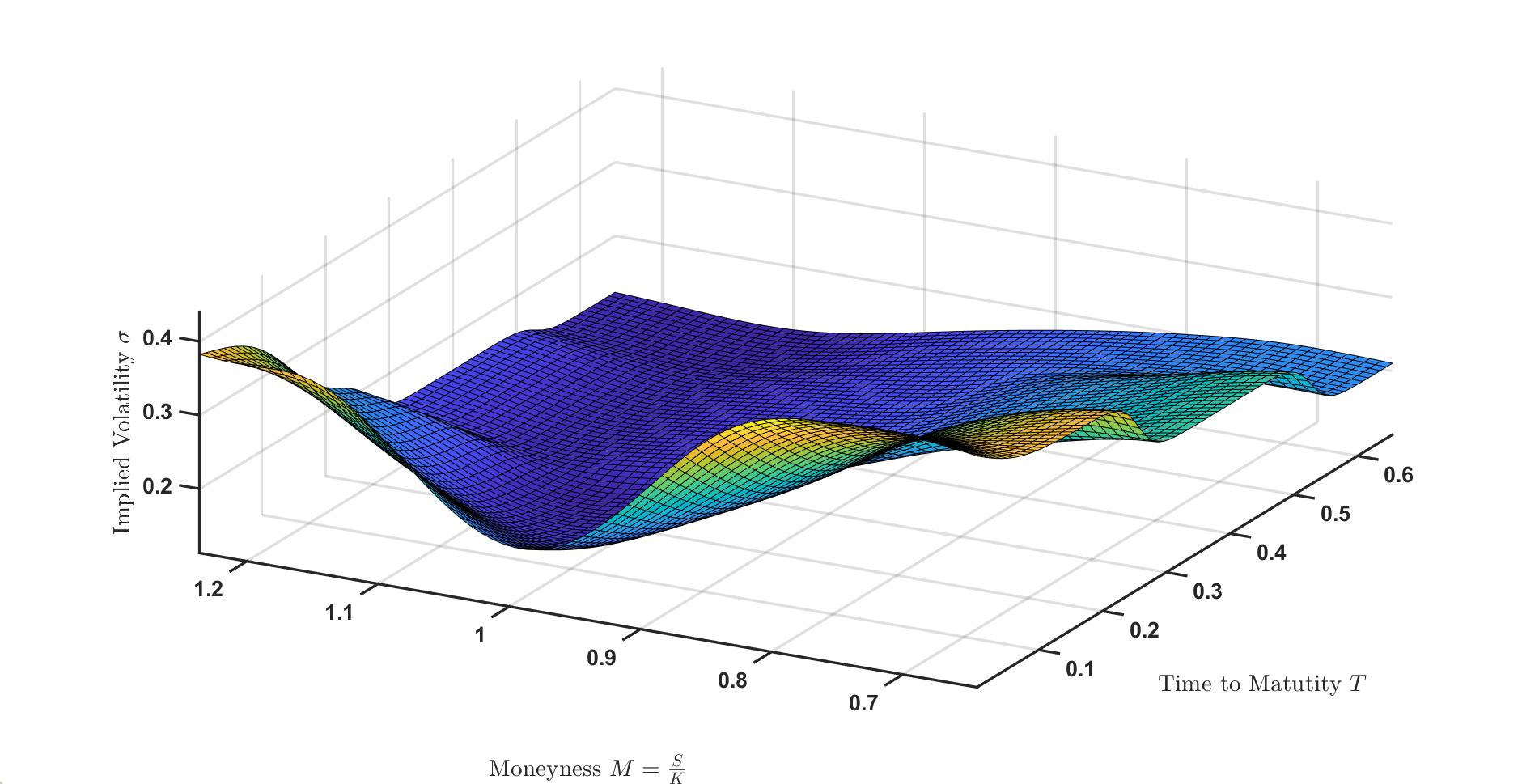}
			\caption{The plot of implied volatility for IVV}
			\label{IV for IVV}
		\end{subfigure}
	\caption{The term-structure of implied volatility of SPY and IVV}
\end{figure}
\begin{figure}
	\centering
		\begin{subfigure}{0.6\textwidth}
			\includegraphics[width=\textwidth]{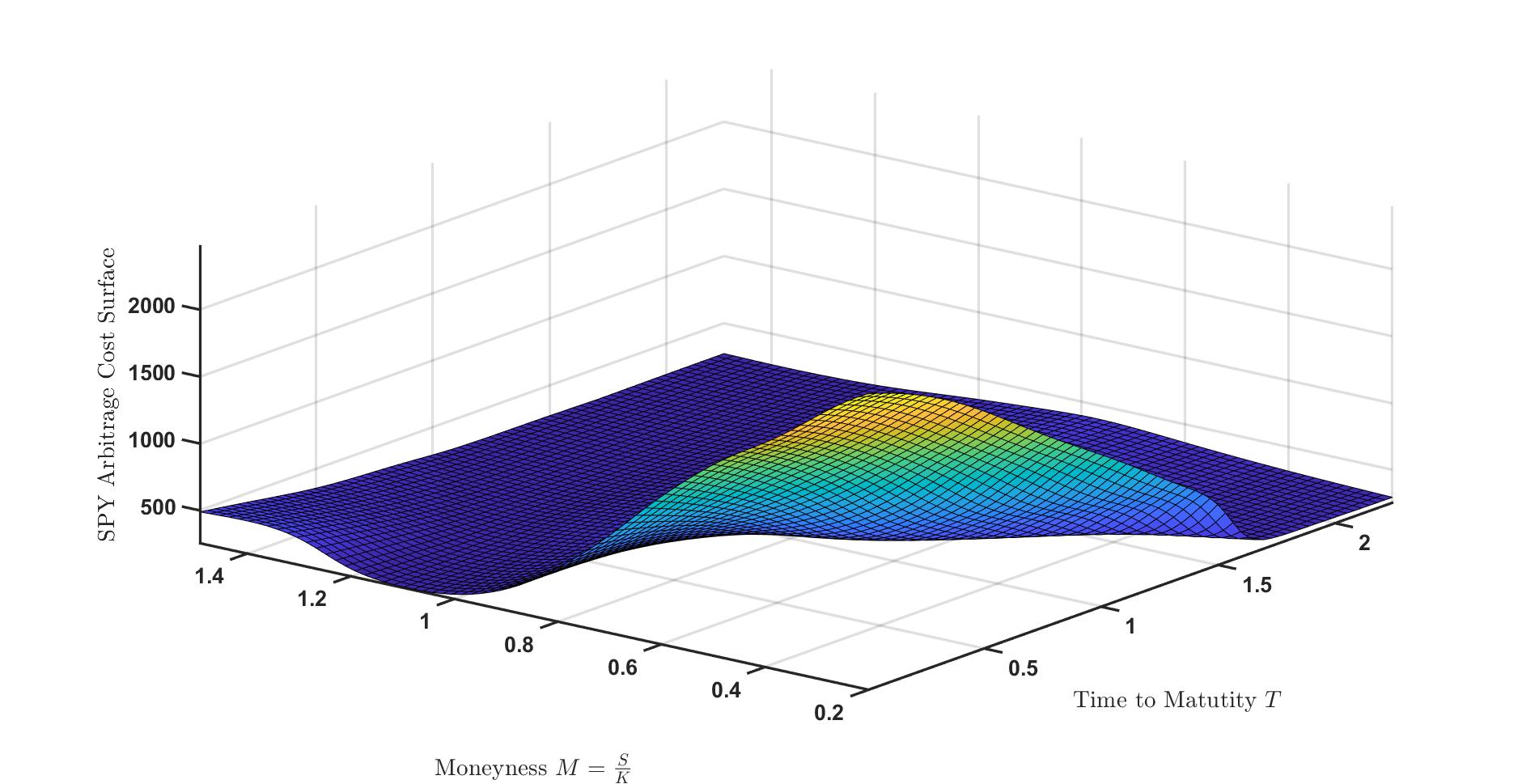}
			\caption{The plot of arbitrage cost surface for SPY}
			\label{ACS for SPY}
		\end{subfigure}
		\begin{subfigure}{0.6\textwidth}
			\includegraphics[width=\textwidth]{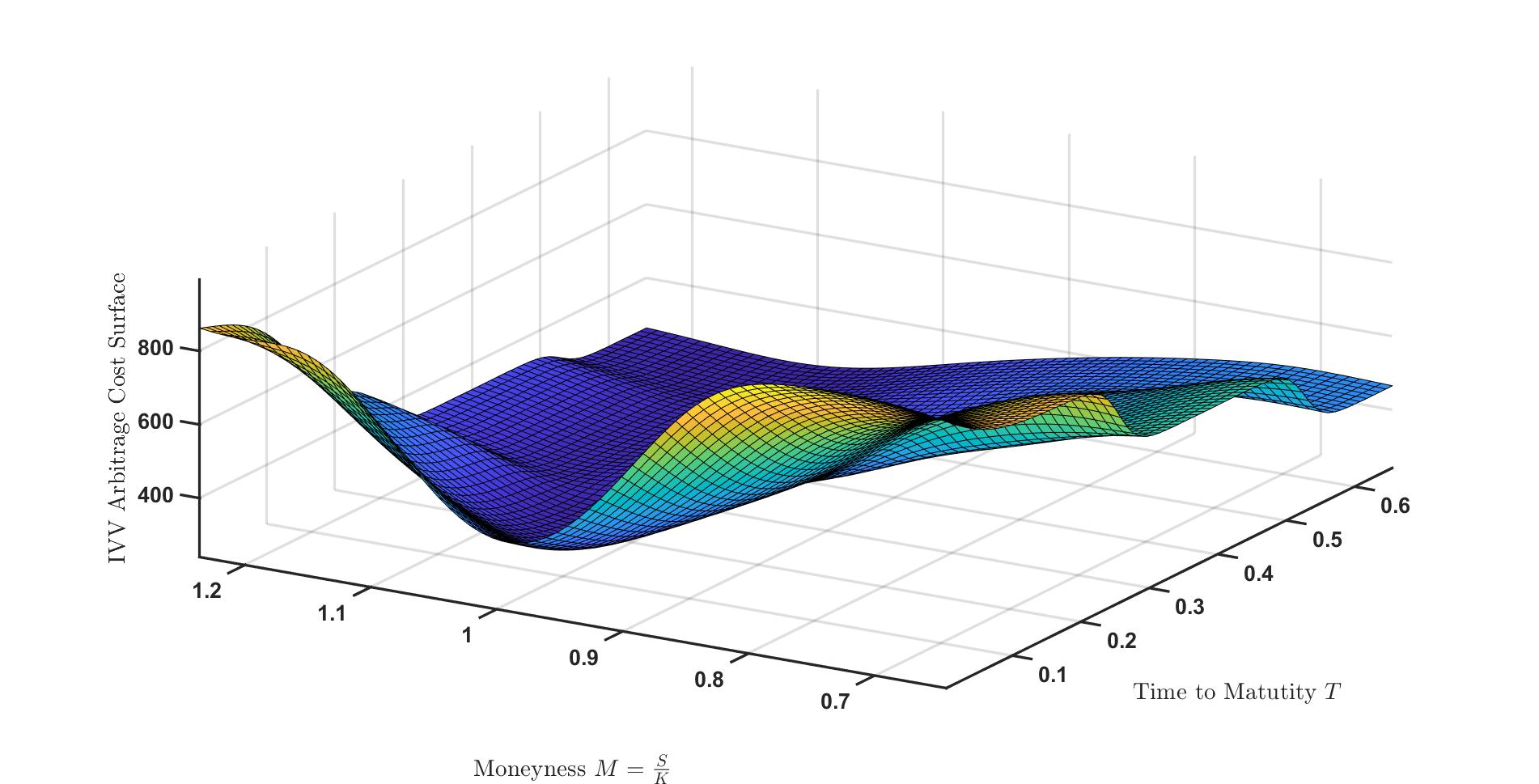}
			\caption{The plot of arbitrage cost surface for IVV}
			\label{ACS for IVV}
		\end{subfigure}
	\caption{The arbitrage cost surfaces of the call options of SPY and IVV}
\end{figure}
\indent We next use the spread prices\footnote{Spread price is equal to ask price minus bid prices.} of options in the construction of the two arbitrage cost surfaces as shown in Figure \ref{spread ACS for SPY} and Figure \ref{spread ACS for IVV} in order to compare the two arbitrage cost surfaces of SPY and IVV. Because of the competitive nature of the market, equilibrium bid-ask spreads should reflect the expected costs of providing liquidity services to the market, and the differences in bid-ask spreads can be directly related to differences in the costs faced by investors (representative investors) across options.\\
\begin{figure}
	\centering
		\begin{subfigure}{0.6\textwidth}
			\includegraphics[width=\textwidth]{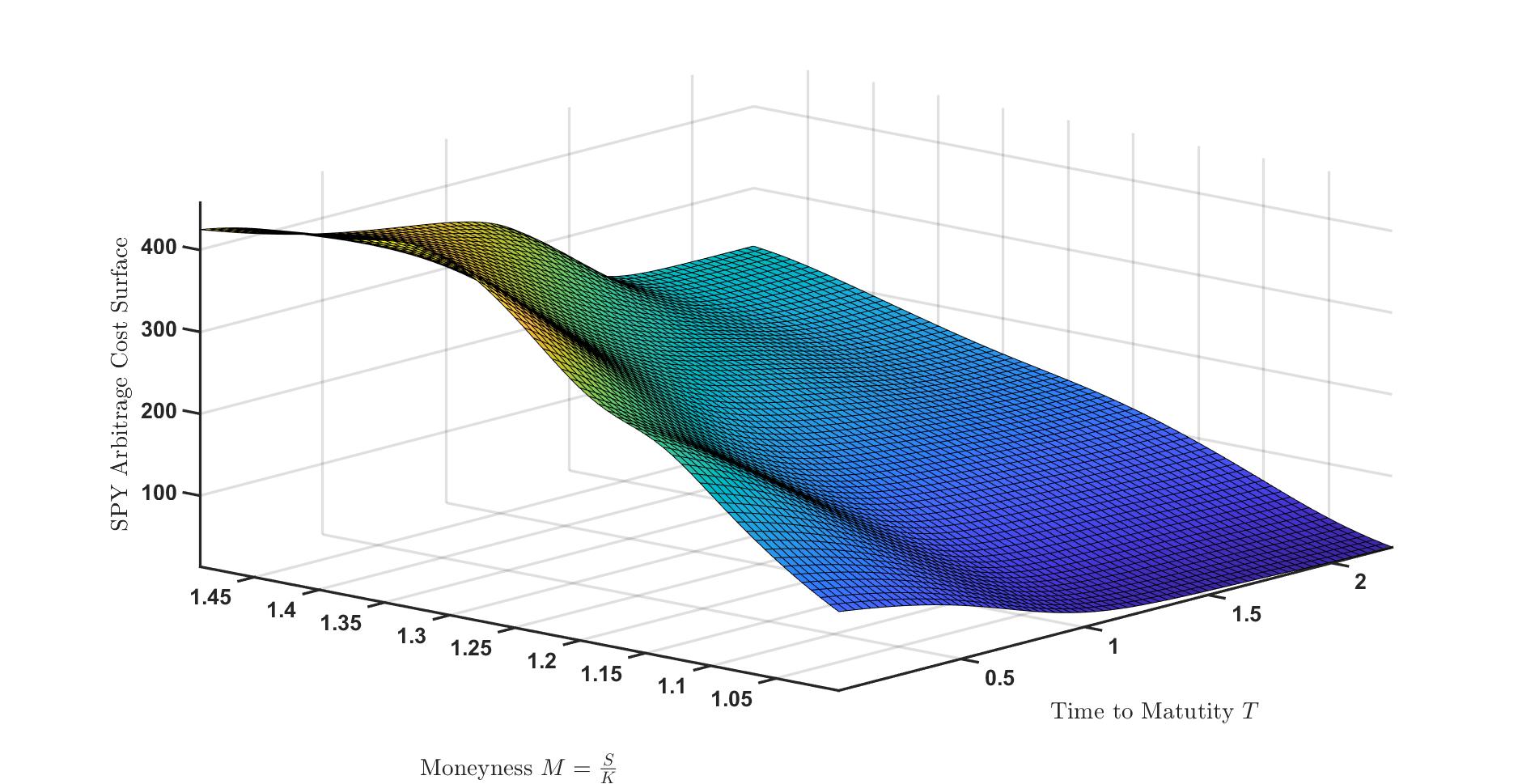}
			\caption{The plot of spread arbitrage cost surface for SPY}
			\label{spread ACS for SPY}
		\end{subfigure}
		\begin{subfigure}{0.6\textwidth}
			\includegraphics[width=\textwidth]{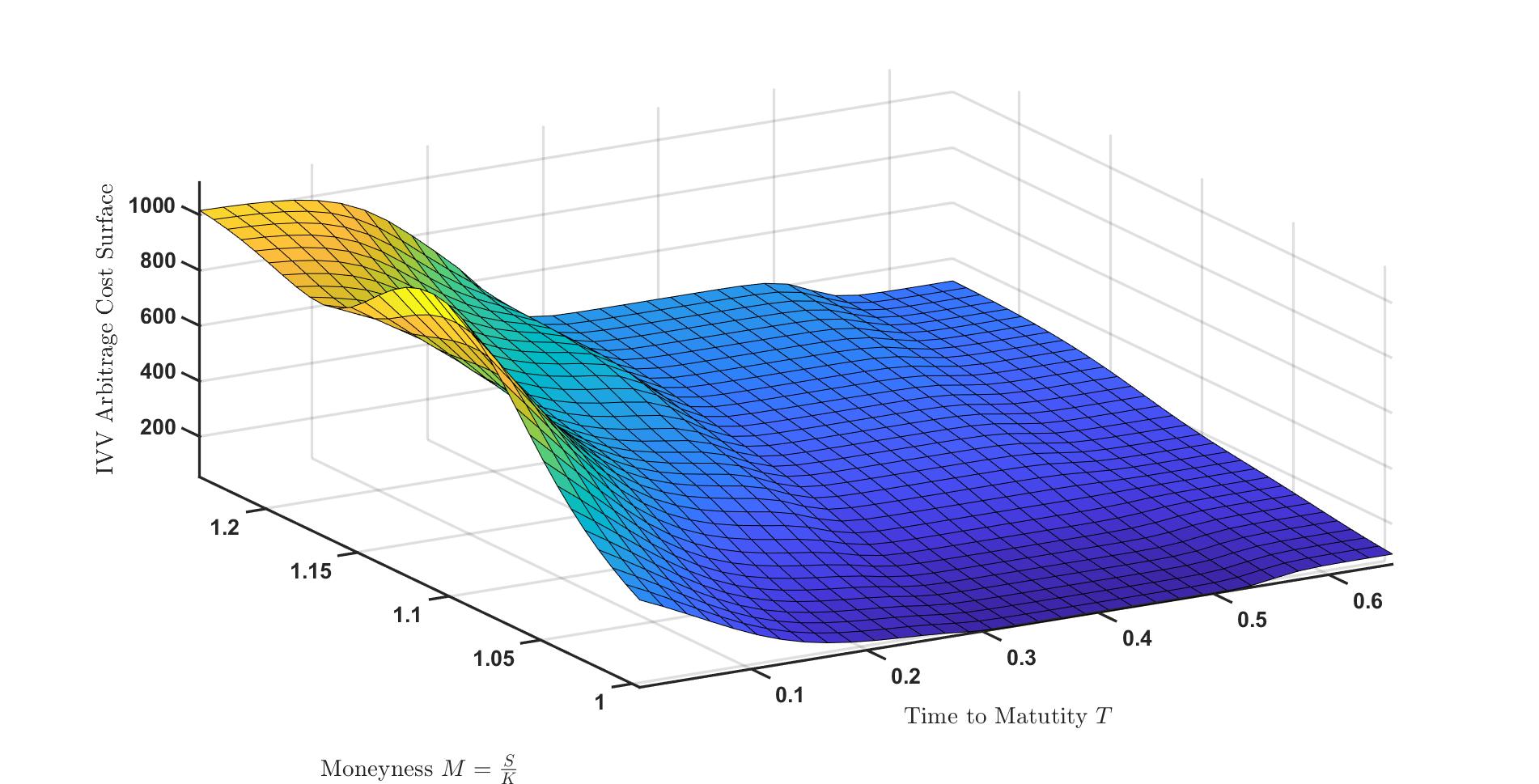}
			\caption{The plot of spread arbitrage cost surface for IVV}
			\label{spread ACS for IVV}
		\end{subfigure}
	\caption{The spread arbitrage cost surfaces of the call options of SPY and IVV}
\end{figure}
\indent In order to visualize any differences of the spread arbitrage cost surfaces, we plot the two surfaces as shown in Figure \ref{spread ACS}. 

As we can observe, the size of the spread arbitrage cost coefficient of the SPY and IVV options in Figure \ref{spread ACS} are highly overlapped and flatten out as the time to maturity increases, which indicate the two market players agreed on the transaction costs when an option contract is put in place. The differs in size of the bid-ask spread arbitrage cost surfaces from SPY to IVV have been observed in higher moneyness with a relative short time-to-maturity of the option contracts, this causes is mainly been considered as the difference in liquidity of each option rather than the difference of the volatility surfaces.\\
\indent Next, the combined arbitrage cost surface for spread prices was plotted by finding the optimized arbitrage cost coefficient value $\mathfrak{C}$ which minimized the price differences of options bid and ask prices and the calibrated Black-Scholes bid and ask prices as shown in Figure \ref{combined ACS}. This combined ACS is determined by the market representative investor ($\mathfrak{N}$, as mentioned in Section 3), whose views are heavily influenced by the market liquidity.
\begin{figure}
\includegraphics[width=\textwidth]{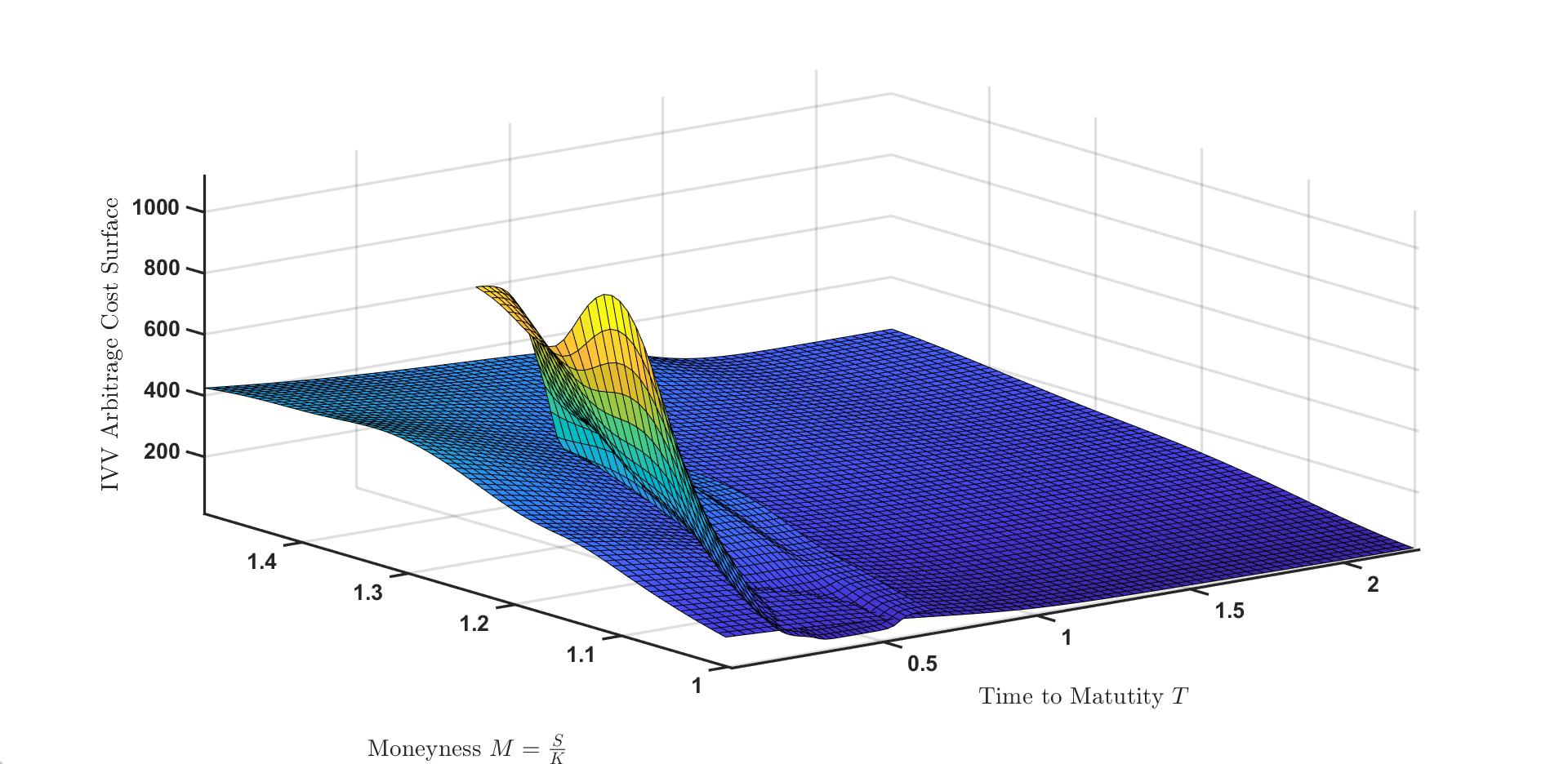} 
	\centering
	\caption{Spread arbitrage cost surface of call options of SPY and IVV}
	\label{spread ACS}
\end{figure}
\begin{figure}[H]
\includegraphics[width=\textwidth]{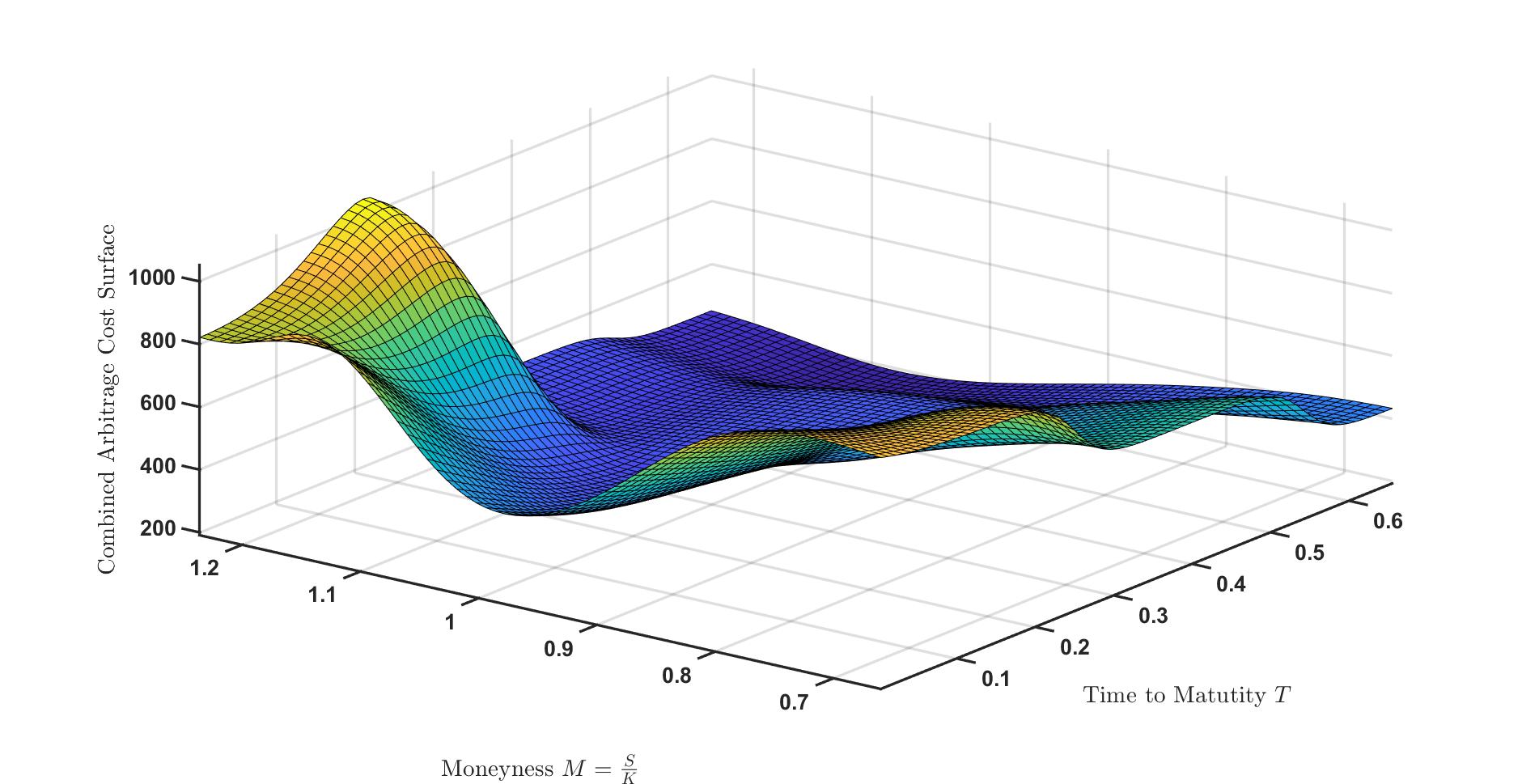} 
	\centering
	\caption{The plot of combined arbitrage cost surface of call options of SPY and IVV}
	\label{combined ACS}
\end{figure}
\section{Conclusion}
In this paper, we correct the statements made by Shefrin (2005) as they pertain to behavioral option pricing. We pointed out that option pricing formulas in Shefrin (2005) has a flaw from the perspective of rational dynamic asset pricing theory. In order to correct Shefrin’s (2005) behavioral approach to option pricing, we introduce arbitrage-costs so that the generated arbitrage gains would be eliminated by arbitrage-costs. We derive the risk-free rate in this setting that generalizes the Black (1972) approach to rational dynamic market with two risky assets driven by the same Brownian motion. In our numerical example, we applied the proposed model to SPY call option prices to calibrate the optimized arbitrage cost coefficient and plotted implied the risk-free rate over time. By comparing the spread bid-ask arbitrage cost surfaces of SPY and IVV call options, we conclude the arbitrage cost coefficient $\mathfrak{C}$ is one and the same for both market participants (SPY and IVV). Lastly, we plot the combined ACS of the market representative investor $\mathfrak{N}$. We believe the effect from belief bias of option traders should be canceled out when the “average” belief is unbiased, and any potential arbitrage gains shall be eliminated by transaction cost.

\end{document}